\documentclass[12pt,journal,onecolumn]{IEEEtran}

\usepackage{cite}
\usepackage{amsmath,amssymb,amsfonts}
\usepackage{graphicx}
\usepackage{textcomp}
\usepackage{xcolor}
\usepackage{algorithm}
\usepackage{algpseudocode}
\usepackage{array}
\usepackage{stfloats}
\usepackage{url}
\usepackage{verbatim}
\usepackage[nice]{nicefrac}
\usepackage{subcaption}
\usepackage[colorlinks,bookmarksopen,bookmarksnumbered,citecolor=red,urlcolor=red]{hyperref}

\def\BibTeX{{\rm B\kern-.05em{\sc i\kern-.025em b}\kern-.08em
    T\kern-.1667em\lower.7ex\hbox{E}\kern-.125emX}}

\begin{document}

\title{Dynamic IRS Allocation for Spectrum-Sharing MIMO Communication and Radar Systems}

\author{Daniyal Munir, Atta Ullah, Danish Mehmood Mughal, \\Min Young Chung, and Hans D. Schotten
}

\maketitle

\begin{abstract}
This paper investigates the use of intelligent reflecting surfaces (IRS) to assist cellular communications and radar sensing operations in a communications and sensing setup. The IRS dynamically allocates reflecting elements to simultaneously localize a target and assist a user's communication. To achieve this, we propose a novel optimization framework that jointly addresses beamforming design and IRS element allocation. Specifically, we formulate a Weighted Minimum Mean Square Error (WMMSE)-based approach that iteratively optimizes the transmit and receive beamforming vectors, IRS phase shifts, and element allocation. The allocation mechanism adaptively balances the number of IRS elements dedicated to communication and sensing subsystems by leveraging the signal-to-noise-plus-interference-ratio (SINR) between the two. The proposed solution ensures efficient resource utilization while maintaining performance trade-offs. Numerical results demonstrate significant improvements in both communication and sensing SINRs under varying system parameters.

\end{abstract}

\begin{IEEEkeywords}
Intelligent reflecting surface (IRS), joint communications and sensing (JCAS), integrated sensing and communications (ISAC), spectrum sharing communication and sensing.
\end{IEEEkeywords}

\section{Introduction}
\label{sec:Intro}
Spectrum has always been the heart of any generation of communication technology and that has not changed for the much-anticipated sixth-generation (6G) of mobile communication \cite{9349624}. Significant efforts are made to efficiently utilize the precious spectrum that faces a continuous threat from the exponential growth of wireless data traffic. In this context, a more focused effort in 6G is directed towards joint communication and sensing (JCAS), where both systems can coexist using the same spectrum and, in some cases, the same hardware \cite{10012421}. This evolution of wireless systems from communication-only networks to dual-functional networks finds applications in the Internet of Things (IoT), vehicle-to-vehicle (V2V), non-terrestrial networks (NTN), nomadic networks, and more \cite{9585321}.

With the popularity of JCAS in 6G, enormous efforts have been dedicated to enabling sensing capabilities, sharing dynamic resources, improving energy efficiency, and co-existing seamlessly (see \cite{10418473} and references therein). Since multi-input-multi-ouput (MIMO) is commonly used in both communication and sensing, beamforming can be intuitively used for spectrum sharing between the two systems \cite{9852716}. The authors in \cite{9838990}, designed a transmit waveform and receive filter for JCAS utilizing the space-time adaptive processing for MIMO radar and symbol-level precoding for MIMO communication. A two-tier alternating optimization framework for spectrum sharing MIMO radar and MIMO communication for peaceful
coexistence is proposed in \cite{8477186}. 

Extending the beamforming design in JCAS, \cite{10042240} deployed an intelligent reflecting surface (IRS) to jointly design a waveform and passive beamform for MIMO JCAS systems. The objective is to maximize the signal-to-noise-plus-interference-ratio (SINR) of the radar and minimize the interference for the communicating users. Further studies have explored IRS-enabled JCAS systems for various objectives, including improving the data rate \cite{9721205}, enhancing localization accuracy \cite{9593143}, and mitigating multi-user interference \cite{10464353}. Another approach proposed the deployment of multiple IRSs to enhance target detection in the absence of a line-of-sight (LoS) path by maximizing the average radar SINR while ensuring a minimal SINR for communication \cite{9743502}. However, these works primarily focus on improving a single parameter, such as data rate or radar accuracy, while providing limited attention to the dynamic trade-offs required in a joint setup. Furthermore, the deployment of multiple IRSs substantially increases both system cost and complexity, limiting their practicality for real-world applications.

In this paper, we present a dynamic IRS-assisted communications and sensing setup, where a single IRS assists the communication of a MIMO base station (BS) and a MIMO radar for their respective operations. The use of a single IRS offers multiple benefits, including reduced deployment costs, smaller radio frequency (RF) footprints, minimized unnecessary interference, and lower signaling overhead. In particular, a multi-antenna BS transmits information symbols to a multi-antenna user while a radar in its proximity detects a moving target, both sharing the same spectrum. The IRS, equipped with multiple reflecting elements, dynamically allocates a subset of its elements for communication and the remainder for sensing. 

In this setup, we propose an IRS-assisted spectrum-sharing framework that iteratively optimizes IRS element allocation, beamforming, and phase shifts to support both communication and sensing operations. Leveraging a Weighted Minimum Mean Square Error (WMMSE)-based approach, the framework enhances the performance of both systems while adaptively balancing element allocation based on their SINR ratios. By dynamically assigning IRS elements to communication and sensing, the proposed framework effectively manages performance trade-offs, mitigates interference, and ensures cost and operational efficiency.

\begin{figure}[t]
    \centering
    \includegraphics[width=0.45\textwidth]{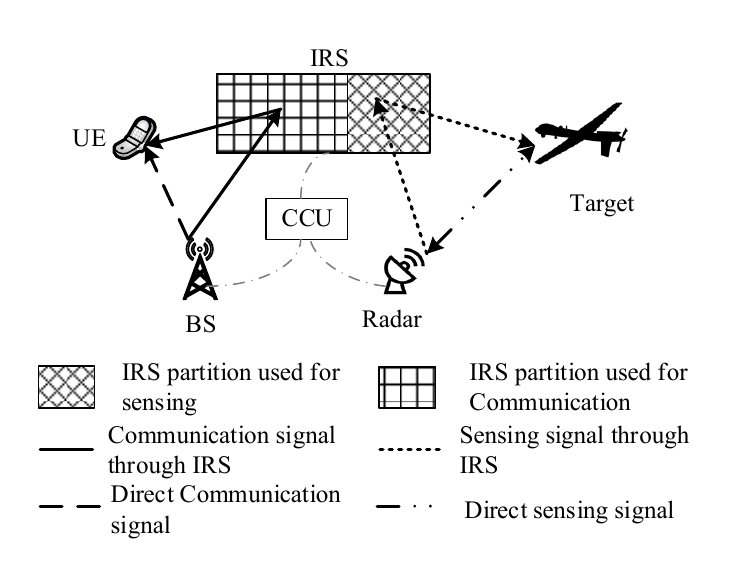}
    \caption{MIMO communication and radar sensing  network setup assisted by a single IRS.}
    \label{fig1:SysMod}
\end{figure}

The rest of the paper is organized as follows. Section \ref{sec:SysMod} introduces the system model for the considered setup and formulates the optimization problem. The proposed solution for element allocation and beamforming design is detailed in Section \ref{sec:PropSch}. The performance evaluation of the proposed optimization framework is presented in Section \ref{sec:Perf}. Finally, the paper is concluded in Section \ref{sec:Con}, along with suggestions for future research directions.

\section{System Model}
\label{sec:SysMod}
\subsection{Network model}
We consider a communication and radar sensing network setup where the BS and radar are deployed as two separate radio units (RUs) in different locations, yet within close proximity to each other. We adopt the same assumptions as described in \cite{8477186}, such as the RUs are connected to the central control unit (CCU) to facilitate coordination between the two RUs. Also, both the systems transmit waveforms at the same symbol period and all the channels are assumed to be known \cite{8477186}. Additionally, an IRS is deployed to simultaneously assist the BS to communicate with a UE and the radar to localize a target. The system model is shown in Fig. \ref{fig1:SysMod}. The BS and the UE have uniform linear arrays (ULAs) with $L$ and $M$ elements, respectively, that are placed at a distance of half a wavelength. The IRS is a uniform planar array (UPA) with $N$ elements that are less than half a wavelength apart. Moreover, $N_c$ elements of the IRS are used for assisting communication while $N_s$ elements are for sensing, such that $N_c + N_s = N$. The radar is deployed with $Q_t$ transmit and $Q_r$ receive antennas, such that, $Q_t = Q_r = Q$ . The radar senses a target present in the network with $K$ reflection points.


\subsection{Channel Model}

For notational simplicity in the subscript, let us denote BS, UE, IRS, radar, and target with $b$, $u$, $i$, $r$, and $t$, respectively. Then, the channel matrices from BS to UE, BS to IRS, BS to radar, radar to UE, radar to IRS, radar to target, IRS to UE, and IRS to target, can be represented by $\mathbf{H}_{b,u} \in \mathbb{C}^{M \times L}$, $\mathbf{H}_{b,i} \in \mathbb{C}^{N \times L}$, $\mathbf{H}_{b,r} \in \mathbb{C}^{Q \times L}$, $\mathbf{H}_{r,u} \in \mathbb{C}^{M \times Q}$, $\mathbf{H}_{r,i} \in \mathbb{C}^{N \times Q}$, $\mathbf{H}_{r,t} \in \mathbb{C}^{K \times Q}$, $\mathbf{H}_{i,u} \in \mathbb{C}^{M \times N}$, and $\mathbf{H}_{i,t} \in \mathbb{C}^{K \times N}$, respectively. We assume that the channel matrices $\mathbf{H}_{b,u}$ and $\mathbf{H}_{r,u}$ follow Rician fading, such that, $\mathbf{H}_{x,y} = \beta_{x,y}\sqrt{\frac{\kappa}{\kappa+1}}\mathbf{H}_{x,y}^\textit{{LoS}}+\beta_{x,y}\sqrt{\frac{1}{\kappa+1}}\mathbf{H}_{x,y}^\textit{{NLoS}}$, where $\beta_{x,y}$ is the distance dependent pathloss, $\kappa$ represents the Rician factor, $\mathbf{H}_{x,y}^\textit{{LoS}}$ is the line-of-sight (LoS) component, and $\mathbf{H}_{x,y}^\textit{{NLoS}}$ is the non-line-of-sight (NLoS) Rayleigh fading components with zero mean and unit variance. For instance, if $\mathbf{H}_{x,y}^\textit{{LoS}}= \mathbf{H}_{b,u}^\textit{{LoS}}$, then $\mathbf{H}_{b,u}^\textit{{LoS}} =  \mathbf{a}_M^{H}(\theta_\text{{UB}})\mathbf{a}_L(\theta_\text{{BU}})$, where $\mathbf{a}_M^{H}(\theta_\text{{UB}})$ is the conjugate transpose of the UE's steering vector pointing towards the BS, and $\mathbf{a}_L(\theta_\text{{BU}})$ is the BS's steering vector pointing towards the UE. In other words, $\theta_\text{{UB}}$ and $\theta_\text{{BU}}$ are the direct-of-arrival (DoA) and direct-of-departeure (DoD), respectively, and the steering vector is given as $\mathbf{a}_M(\theta_\text{{UB}}) \triangleq  [1, e^{-j\pi\sin \theta_\text{UB}}, \cdots, e^{-j (M-1)\pi\sin \theta_\text{UB}}]^H$. 

It is rational to assume that all other channels have a LoS component since the IRS should be deployed to have a LoS connection with both the BS and radar to assist them, and we assume that the target is airborne. In addition, the distance dependant pathloss $\beta_{x,y}$ is modeled as $\beta_{x,y} = C_0 ({D_{x,y}}/{D_0})^{-\alpha}$, where $D_{x,y}$ is the propagation distance between $x$ and $y$, $C_0 = (\lambda/4\pi D_0)^2$ is the free space path loss at a reference distance $D_0 = 1$m, $\lambda$ is the wavelength, and $\alpha$ represents the pathloss exponent.

\subsection{Signal Model}
For an IRS-assisted JCAS system, there are too many radio footprints in the environment that can interfere with each other. For sensing, the radar-target-radar signal has multiple paths, the direct path radar-target-radar, radar-IRS-target-radar, radar-target-IRS-radar, and the radar-IRS-target-IRS-radar. Since the target does not use beamforming (either active or passive), we assume that the doubly reflected signal has weakened and would not significantly affect other signals. This assumption is reasonable as these doubly reflected signals can be treated as just two additional multipaths. Therefore, the last two paths mentioned above are neither included for quantifying signal strength for sensing nor for the interference at the UE.  

Subsequently, the received signal at the UE can be given as 
\begin{equation} 
\label{RS_UE}
\begin{aligned}
&\mathbf{y}_c = \mathbf{w}_u^H (\mathbf{H}_{b,u} + \mathbf{H}_{i,u} \mathbf{\Phi}^c \mathbf{H}_{b,i}) \mathbf{w}_b s_b + 
\\&
\mathbf{w}_u^H (\mathbf{H}_{r,u} +  \mathbf{H}_{i,u} \mathbf{\Phi}^s \mathbf{H}_{r,i} + \mathbf{H}_{t,u}\rho\mathbf{H}_{r,t}) \mathbf{w}_r s_r + \mathbf{w}_u^H\mathbf{n}_u,
\end{aligned}
\end{equation}
where $\mathbf{w}_b$ and $\mathbf{w}_r$ are the transmit beamforming vectors at the BS and radar, respectively, and $\mathbf{w}_u$ is the receive beamforming vector at the UE, $\rho$ is the reflection coefficient of the target, $\mathbf{\Phi}^c = \text{diag}(\phi_1^c, \phi_2^c,\ldots, \phi_{N_c}^c)$, $\mathbf{\Phi}^s = \text{diag}(\phi_{N_c + 1}^s, \phi_{N_c + 2}^s,\ldots, \phi_N^s)$ are the respective phase shift matrices of the IRS elements assisting communication and sensing. Moreover, $s_b$ is the transmitted signal from the BS, $s_r$ is the transmitted signal from the radar, and $\mathbf{n}_u$ is the noise vector at the UE, modeled as $\mathcal{CN}(0, \sigma_u^2 \mathbf{I})$. For the sake of simplicity, let us denote 
$\mathbf{H}_{des}^c = \mathbf{H}_{b,u} + \mathbf{H}_{i,u} \mathbf{\Phi}^c \mathbf{H}_{b,i}$ as a desired channel for communication and 
$\mathbf{H}_{int}^c = \mathbf{H}_{r,u} + \mathbf{H}_{i,u} \mathbf{\Phi}^s \mathbf{H}_{r,i} + \mathbf{H}_{t,u}\rho\mathbf{H}_{r,t}$ as the interfering channel for communication. The corresponding SINR can be given as 
\begin{equation}
\label{SINR_UE}
\begin{aligned}
\gamma_{c} = \frac{\left\| \mathbf{w}_u^H \mathbf{H}_{des}^c\mathbf{w}_b\right\|^2}{\left\|\mathbf{w}_u^H \mathbf{H}_{int}^c\mathbf{w}_r\right\|^2 + \mathbf{w}_u^H \sigma_u^2},
\end{aligned}
\end{equation}
where $\sigma_{u}^2$ is the noise variance at the UE.

Similarly, for sensing, the received signal at the radar, reflected from the target, and the interfering signals from BS and IRS can be given as
\begin{equation}
\label{RS_R}
\begin{aligned}
&\mathbf{y}_s \!\!=\!\! \tilde{\mathbf{w}}_r^H (\mathbf{H}_{t,r}\rho\mathbf{H}_{r,t}\!\! +\!\!   \mathbf{H}_{t,r} \rho\mathbf{H}_{i,t} \mathbf{\Phi}^s \mathbf{H}_{r,i}) \mathbf{w}_r s_r  \!\!+ \!\!
\tilde{\mathbf{w}}_r^H (\mathbf{H}_{b,r} \! + 
\\&
\mathbf{H}_{i,r} \mathbf{\Phi}^c \mathbf{H}_{b,i} \!  \!+\!\! \rho  \mathbf{H}_{t,r}  \mathbf{H}_{b,t} \!\! + \!\! \rho  \mathbf{H}_{t,r} \mathbf{H}_{i,t} \mathbf{\Phi}^c \mathbf{H}_{b,i}) \mathbf{w}_b s_b \!+\! \tilde{\mathbf{w}}_r^H\mathbf{n}_r
\end{aligned}
\end{equation}
where $\tilde{\mathbf{w}}_r$ is the receive beamforming vector at the radar, $\mathbf{n}_r$ is the noise vector at the radar, modeled as $\mathcal{CN}(0, \sigma_r^2 \mathbf{I})$. To simplify the subsequent SINR at the radar, let us denote 
$\mathbf{H}_{des}^s =\mathbf{H}_{t,r}\rho\mathbf{H}_{r,t} +   \mathbf{H}_{t,r} \rho\mathbf{H}_{i,t} \mathbf{\Phi}^s \mathbf{H}_{r,i}$ as a desired channel for sensing and $\mathbf{H}_{int}^s=\mathbf{H}_{b,r}  + \mathbf{H}_{i,r} \mathbf{\Phi}^c \mathbf{H}_{b,i} + \rho  \mathbf{H}_{t,r}  \mathbf{H}_{b,t} + \rho  \mathbf{H}_{t,r} \mathbf{H}_{i,t} \mathbf{\Phi}^c \mathbf{H}_{b,i}$ as the interfering channel for sening, which is given as  
\begin{equation}
\label{SINR_R}
\begin{aligned}
\gamma_s = \frac{\left\|\tilde{\mathbf{w}}_r^H \mathbf{H}_{des}^s \mathbf{w}_r\right\|^2}{\left\|\tilde{\mathbf{w}}_r^H\mathbf{H}_{int}^s\mathbf{w}_b \right\|^2 + \tilde{\mathbf{w}}_r^H\sigma_r^2},
\end{aligned}
\end{equation}
where $\sigma_r^2$ is the noise power at the radar. Different waveform designs or signal processing techniques can be employed to mitigate the effect of interference \cite{10012421}, \cite{9627227}, however, our main aim in this work is to maximize the SINR when an IRS is employed in a communication and sensing network setup in the presence of interference.

\subsection{Problem Formulation}
To design beamforming vectors and phase shifts that maximize the SINRs for both systems while accounting for cross-system interference, we adopt a Weighted Minimum Mean Square Error (WMMSE) approach. The WMMSE framework is commonly used to reformulate the SINR constraints and optimize the beamforming vectors. 
Respective SINRs can be expressed in terms of MSE as given below,
\begin{equation}
\label{MSE}
\begin{aligned}
\Gamma_c = \mathbb{E}[|\hat{s_b}-s_b|^2] \quad \text{and} \quad \Gamma_s = \mathbb{E}[|\hat{s_r}-s_r|^2],
\end{aligned}
\end{equation}
where $\Gamma_c$ and $\Gamma_s$ represent MSE at the UE and radar, respectively, $\hat{s_b}= \mathbf{w}^H_u\mathbf{y}_c$ and $\hat{s_r}= \tilde{\mathbf{w}}_r^H\mathbf{y}_s$ are the estimated symbols for communication and sensing, respectively. The problem can be formulated as a joint optimization problem, aiming to find the optimal $\mathbf{w_b}$, $\mathbf{w_r}$, $\mathbf{\Phi}^c$, $\mathbf{\Phi}^s$, $N_c$, and $N_s$ that minimizes the weighted MSEs $\Gamma_c$ and $\Gamma_s$. This can be represented as:
\begin{equation}
\label{Opt}
\begin{aligned}
\min_{\mathbf{w_b}, \mathbf{w_r}, \mathbf{\Phi}, N} \eta \Gamma_c + (1-\eta)\Gamma_s\\
\text{subject to: } N_c + N_s = N,\\ 
0\leq \phi_n^c, \phi_n^s \leq 2\pi,  & \quad \forall n, \\
\left\|\mathbf{w_b}\right\|^2 \leq P^c_{max},
\left\|\mathbf{w_r}\right\|^2 \leq P^s_{max},
\end{aligned}
\end{equation}
where $\eta = \frac{1}{1 + \left( \gamma_c/\gamma_s \right)^\beta}$ is the dynamic element allocation weight, such that, $N_c = N \cdot \eta$, and $N_s = N \cdot (1 - \eta)$. The parameter $\beta$ controls the sensitivity of the SINR ratio, such that, a small value of $\beta$ gradually changes the allocation of elements while a large value switches the allocation sharply between communication and sensing. In addition, the first constraint defines that all the elements of the IRS should be allocated, either for communication or sensing. The second constraint limits the range of phase shifts between 0 and 2$\pi$. 
The last constraint is included to limit the transmit power of the BS and radar to their respective maximum allowable transmit power $P^c_{max}$ and $P^s_{max}$.

The optimization problem (\ref{Opt}) is highly non-convex due to the multiplication of variables such as the reflection coefficients $\mathbf{\Phi}^c$ and $\mathbf{\Phi}^s$, as well as beamforming vectors $\mathbf{w_b}$ and $\mathbf{w_r}$. The presence of bilinear terms, where IRS elements are multiplied by the beamforming vectors, further complicates the problem.  Furthermore, the optimization variables are coupled through constraints such as SINR constraints for both UE and radar and power constraints. This coupling makes it difficult to separate variables and solve this non-convex problem. Since IRS elements influence both communication and sensing, a joint optimization that balances these objectives requires iterative numerical methods.

\section{Element Allocation and Beamforming Design}
\label{sec:PropSch}
In this section, we present an IRS-assisted spectrum-sharing framework that solves this highly non-convex optimization problem. Specifically, we adopt an iterative block coordinate descent (BCD) approach, with the WMMSE framework as the underlying method for optimizing beamforming vectors and phase shifts. To make the problem tractable, it is decomposed into smaller, more manageable subproblems, and iteratively optimize beamforming vector $\mathbf{w}_b$, $\mathbf{w}_r$, $\mathbf{w}_u$, and $\tilde{\mathbf{w}}_r$, phase shifts $\mathbf{\Phi}^c$ and $\mathbf{\Phi}^s$, and element allocation $N_c$ and $N_s$.

\subsection{WMMSE-Based Joint Optimization}

The main WMMSE-based joint optimization algorithm alternates between optimizing the transmit and receive beamforming vectors and phase shifts optimization and element allocation for communication and sensing systems. The iterative BCD-based optimization procedures used to obtain the optimal values for the MIMO communication and radar systems are summarized in Algorithm \ref{algo1:MJOA}.

Based on random initialization of beamforming vectors and phase shifts, the SINRs of both the systems are computed using (\ref{SINR_UE}) and (\ref{SINR_R}). The ratio of SINRs $\gamma_c$ and $\gamma_s$ and weighting factor $\beta$ are then used to calculate the element allocation parameters $\eta$, $N_c$, and $N_s$. The convergence tolerance $\epsilon$ and iteration counter $k$ are also set. After the initialization of all the required parameters, the optimization process starts where beamforming vectors and phase shifts are first optimized using the steps summarized in algorithm \ref{algo2:BFPS}. The optimized beamforming vectors and phase shifts received from Algorithm \ref{algo2:BFPS}, are used to recompute the SINRs $\gamma_c$ and $\gamma_s$. Subsequently, element allocation parameters are updated to allocate IRS elements optimally for each system. The objective function value $f^{(k)} = \eta \Gamma_c + (1 - \eta) \Gamma_s$ is then evaluated and if the change, $|f^{(k)} - f^{(k-1)}|$, is less than the $\epsilon$, the algorithm terminates. These steps are carried out iteratively until a convergence criterion is met. 

\begin{algorithm}[h]
\caption{WMMSE-Based Joint Optimization Algorithm}
\label{algo1:MJOA}
\begin{algorithmic}[1]
\State \textbf{Initialization:}
\State Initialize transmit beamforming vectors $\mathbf{w}_b$, $\mathbf{w}_r$, and receive beamforming vectors $\mathbf{w}_u$, $\tilde{\mathbf{w}}_r$.
\State Initialize random phase shift matrices $\mathbf{\Phi}^c$ and $\mathbf{\Phi}^s$.
\State Set convergence tolerance $\epsilon > 0$.
\State Compute initial $\eta = \frac{1}{1 + \left( \frac{\gamma_c}{\gamma_s} \right)^\beta}$, $N_c = N \cdot \eta$, and $N_s = N \cdot (1 - \eta)$.
\State Set iteration counter $k = 0$.

\Repeat
    \State \textbf{Beamforming Vector Optimization}
    \State Call Algorithm 2 (Beamforming Vector and Phase Shift Optimization) to solve for $\mathbf{w}_b$, $\mathbf{w}_r$, $\mathbf{w}_u$, $\tilde{\mathbf{w}}_r$, $\mathbf{\Phi}^c$, and $\mathbf{\Phi}^s$.

    \State \textbf{Update $\eta$ and Element Allocation}
    \State Compute $\gamma_c$ and $\gamma_s$ with optimized parameters

    \State Recalculate $\eta = \frac{1}{1 + \left( \frac{\gamma_c}{\gamma_s} \right)^\beta}$.
    \State Update $N_c = N \cdot \eta$ and $N_s = N \cdot (1 - \eta)$.

    \State \textbf{Check Convergence}
    \State Compute the objective function value $f^{(k)} = \eta \Gamma_c + (1 - \eta) \Gamma_s$.
    \State If $|f^{(k)} - f^{(k-1)}| \leq \epsilon$, stop. Otherwise, increment $k$ and repeat.

\Until convergence

\State \textbf{Output:} Optimal $\mathbf{w}_b$, $\mathbf{w}_r$, $\mathbf{w}_u$, $\tilde{\mathbf{w}}_r$, $\mathbf{\Phi}^c$, $\mathbf{\Phi}^s$, $\eta$, $N_c$, and $N_s$.

\end{algorithmic}
\end{algorithm}

\subsection{Beamforming and Phase Shift Optimization}

The first step in this optimization process is to design the receive beamforming vectors for UE $\mathbf{w}_u$ and radar receive antennas $\tilde{\mathbf{w}}_r$. To keep the design of $\mathbf{w}_u$ and $\tilde{\mathbf{w}}_r$ simple, we decompose the respective desired channels $\mathbf{H}_{\text{des}}^c$ and $\mathbf{H}_{\text{des}}^s$. 
Performing singular value decomposition (SVD) allows the decomposition of the desired channel matrix into orthogonal eigenmodes. By aligning $\mathbf{w}_u$ and $\tilde{\mathbf{w}}_r$ with the dominant left singular vector of the $\mathbf{H}_{\text{des}}^c$ and $\mathbf{H}_{\text{des}}^s$, respectively, the received signal strength for both communication and sensing systems can be maximized. Using SVD reduces the complexity of designing the receive beamformer optimization by decoupling it from interference management, which is duly considered in the regularization-based transmit beamformer design.

Designing optimal transmit beamforming vectors in this spectrum-sharing scenario is challenging because of the cross-interference caused by both systems. In addition, power constraints limit the magnitude of these beamforming vectors. To address these challenges, the transmit beamformers are designed using a regularization-based approach that explicitly accounts for interference \cite{10742291}. The regularization parameter ($\mu$) ensures that the transmitted signals are optimized not only for maximizing the desired signal but also for mitigating the interference. $\mu$ is defined as the ratio of the interference channel power to the desired channel power,
\begin{equation}
\label{MU}
\begin{aligned}
\mu = \frac{\text{Tr}(\mathbf{H}_{\text{int}} \mathbf{H}_{\text{int}}^{H})}{\text{Tr}(\mathbf{H}_{\text{des}} \mathbf{H}_{\text{des}}^{H})}.
\end{aligned}
\end{equation}
This dynamic selection of $\mu$ offers a balance between interference suppression and maximizing the desired signal. This regularization-based approach is often referred to as regularized zero-forcing or MMSE beamforming \cite{6832894}. Subsequently, the transmit beamforming vectors can be given as  
\begin{equation}
\label{OptBF}
\begin{aligned}
\mathbf{w}_b = (\mathbf{H}_{\text{des}}^{c H} \mathbf{w}_u \mathbf{w}_u^H \mathbf{H}_{\text{des}}^c + \mu_c \mathbf{I})^{-1} \mathbf{H}_{\text{des}}^{c H} \mathbf{w}_u,\\
\mathbf{w}_r = (\mathbf{H}_{\text{des}}^{s H} \tilde{\mathbf{w}}_r \tilde{\mathbf{w}}_r^H \mathbf{H}_{\text{des}}^s + \mu_s \mathbf{I})^{-1} \mathbf{H}_{\text{des}}^{s H} \tilde{\mathbf{w}}_r,
\end{aligned}
\end{equation}
\begin{algorithm}[h]
\caption{Beamforming Vector and Phase Shift Optimization with Regularization}
\label{algo2:BFPS}
\begin{algorithmic}[1]
\State \textbf{Input:} Current $\mathbf{w}_b$, $\mathbf{w}_r$, $\mathbf{w}_u$, $\tilde{\mathbf{w}}_r$, $\mathbf{\Phi}^c$, $\mathbf{\Phi}^s$.
\State \textbf{Update Receive Beamforming Vectors}
\State Compute $\mathbf{w}_u$ and $\tilde{\mathbf{w}}_r$ by performing SVD on the respective desired channels
\State \textbf{Update Transmit Beamforming Vectors with Regularization}
\State Compute $\mu$ as the ratio of channel powers
\State Compute $\mathbf{w}_b$ and $\mathbf{w}_r$ using regularization:
\[ \mathbf{w}_b = (\mathbf{H}_{\text{des}}^{c H} \mathbf{w}_u \mathbf{w}_u^H \mathbf{H}_{\text{des}}^c + \mu_c \mathbf{I})^{-1} \mathbf{H}_{\text{des}}^{c H} \mathbf{w}_u, \]
\[ \mathbf{w}_r = (\mathbf{H}_{\text{des}}^{s H} \tilde{\mathbf{w}}_r \tilde{\mathbf{w}}_r^H \mathbf{H}_{\text{des}}^s + \mu_s \mathbf{I})^{-1} \mathbf{H}_{\text{des}}^{s H} \tilde{\mathbf{w}}_r. \]
\State \textbf{Update Phase Shifts for Communication and Sensing}
\State  \textbf{Communication Phase Shifts:}
\For{$n = 1$ to $N_c$}
    \State Compute $\phi_n^c = \arg\left((\mathbf{H}_{i,u})_n^H \mathbf{H}_{b,i} \mathbf{w}_b \mathbf{w}_u^H\right)$.
\EndFor
\State  \textbf{Sensing Phase Shifts:}
\For{$n = 1$ to $N_s$}
    \State Compute $\phi_n^s = \arg\left((\mathbf{H}_{i,t})_n^H \mathbf{H}_{r,i} \mathbf{w}_r \tilde{\mathbf{w}}_r^H\right)$.
\EndFor
\State Construct $\mathbf{\Phi}^c = \text{diag}(e^{j\phi_1^c}, \ldots, e^{j\phi_{N_c}^c})$ and $\mathbf{\Phi}^s = \text{diag}(e^{j\phi_1^s}, \ldots, e^{j\phi_{N_s}^s})$.
\State \textbf{Output:} Updated $\mathbf{w}_b$, $\mathbf{w}_r$, $\mathbf{w}_u$, $\tilde{\mathbf{w}}_r$, $\mathbf{\Phi}^c$, and $\mathbf{\Phi}^s$.
\end{algorithmic}
\end{algorithm}
where $\mu_c$ and $\mu_s$ can be calculated with $\mathbf{H}_{\text{int}}^c$, $\mathbf{H}_{\text{des}}^c$ and $\mathbf{H}_{\text{int}}^s$, $\mathbf{H}_{\text{des}}^s$, respectively. Finally, beamforming vectors are normalized to satisfy the power constraints given in (\ref{Opt}). The regularization parameter along with the identity matrix also offers numerical stability and robustness to channel uncertainty \cite{6832894}.

These optimized beamforming vectors are then used to compute the optimal phase shifts $\mathbf{\Phi}^c$ and $\mathbf{\Phi}^s$. Intuitively, to maximize the SINR for communication, the phase shift $\phi_n^c$ is chosen such that the reflected signal constructively combines with the direct signal. Since we assume that all the channels are known, the incident signals at the IRS can be reflected in the desired direction by adjusting its phase shift. For each reflecting element $n$, $\forall n \in \{1,2, \cdots, N_c\}$, the phase shift is adjusted to align the IRS-reflected path with the desired signal direction and can be  given as,
\begin{equation}
\label{phic}
\begin{aligned}
\phi_n^c = \arg\left((\mathbf{H}_{i,u})_n^H \mathbf{H}_{b,i} \mathbf{w}_b \mathbf{w}_u^H\right).
\end{aligned}
\end{equation}
Using this updated value of each reflecting element, $\mathbf{\Phi}^c$ is reconstructed. Similarly, the phase shift for sensing, $\phi_n^s$, $\forall n \in \{1,2, \cdots, N_s\}$, are calculated as,
\begin{equation}
\label{phic}
\begin{aligned}
\phi_n^s = \arg\left((\mathbf{H}_{i,t})_n^H \mathbf{H}_{r,i} \mathbf{w}_r \tilde{\mathbf{w}}_r^H\right).
\end{aligned}
\end{equation}
Finally, the updated values are used to reconstruct $\mathbf{\Phi}^s$. The steps to design the beamforming vector and phase shifts are summarized in Algorithm (\ref{algo2:BFPS}). The updated values of beamforming vectors and phase shifts are then forwarded to Algorithm (\ref{algo1:MJOA}) for further operations.

\section{Performance Analysis}
\label{sec:Perf}
In this section, we evaluate the performance of the proposed IRS-assisted spectrum-sharing framework. In the simulation setup, the BS is placed at the origin of a 2D Cartesian coordinate system, and the UE is randomly deployed within the radius of the BS. A radar is also deployed randomly in the proximity of the BS so that it also lies within its radius. A target is then deployed within the radius of the radar. The IRS is positioned at a fixed location between the BS and the radar to ensure efficient signal reflection for both systems. The channels between the network entities are generated randomly for each simulation run based on their respective fading models. Unless stated otherwise, all the simulation parameters used in the simulations are given in Table \ref{tbl:SP}. The simulation runs are averaged over 5000 random channel realizations.

\begin{table}[h]
\centering
\caption{Simulation Parameters}
\label{tbl:SP}
\begin{tabular}{|l|l|l|}
\hline
\textbf{Parameter} & \textbf{Symbol} & \textbf{Value} \\ \hline
BS radius & $R_b$ & 200 m \\ \hline
Radar radius & $R_r$ & 200 m \\ \hline
No. of antennas at BS & $L$ & 8 \\ \hline
No. of antennas at UE & $M$ & 2 \\ \hline
No. of antennas at Radar & $Q$ & 8 \\ \hline
No. of IRS elements & $N$ & 32 \\ \hline
No. of reflecting points of a target & $K$ & Random [1-5] \\ \hline
Reflection coefficient of target & $\rho$ & 0.2 \\ \hline
Operating frequency & $f$ & 31 GHz \\ \hline
Pathloss exponent & $\alpha$ & 3.5 \\ \hline
Maximum BS transmit power & $P^c_{\text{max}}$ & 10 W \\ \hline
Maximum Radar transmit power & $P^s_{\text{max}}$ & 10 W \\ \hline
Convergence threshold & $\epsilon$ & \(1 \times 10^{-3}\) \\ \hline
Balancing factor & $\beta$ & 0.5 \\ \hline
\end{tabular}
\label{tab:simulation_parameters}
\end{table}

\begin{figure}[h]
    \centering
    \includegraphics[width=0.42\textwidth]{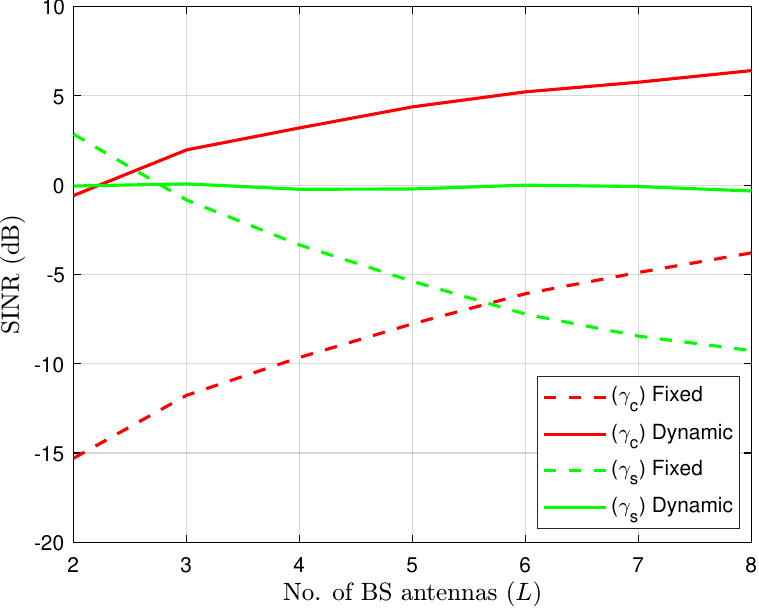}
    \caption{Comparison of fixed and dynamic IRS element allocation.}
    \label{fig2:SINR}
\end{figure}

The objective of the proposed optimization framework is to enhance the SINR of both systems. Fig. \ref{fig2:SINR} shows the SINR values for the varying number of BS antennas for different allocations of IRS elements. The proposed dynamic allocation clearly shows improvement in SINR values for both communication and sensing as compared to fixed allocation ($N_c=N_s=N/2$). For fixed allocation, the sensing SINR decreases as $L$ increases due to the rising interference from the increasing number of BS antennas. However, in the proposed scheme, IRS elements are dynamically reallocated for sensing, effectively mitigating the impact of increased interference. This adaptive adjustment prevents the sensing SINR from decreasing, ensuring stable performance despite the additional interference. The fixed allocation not only represents the division of elements within a single IRS but also serves as a practical example of a scenario where separate IRSs are deployed to assist both communication and sensing. In such a case, the allocated $N_c$ and $N_s$ elements would correspond to the sizes of the co-located IRSs dedicated to communication and sensing, respectively. This highlights the effectiveness of the dynamic allocation strategy over fixed allocation and scenarios similar to those considered in \cite{9743502}. 

\begin{figure}[h]
    \centering
    \includegraphics[width=0.42\textwidth]{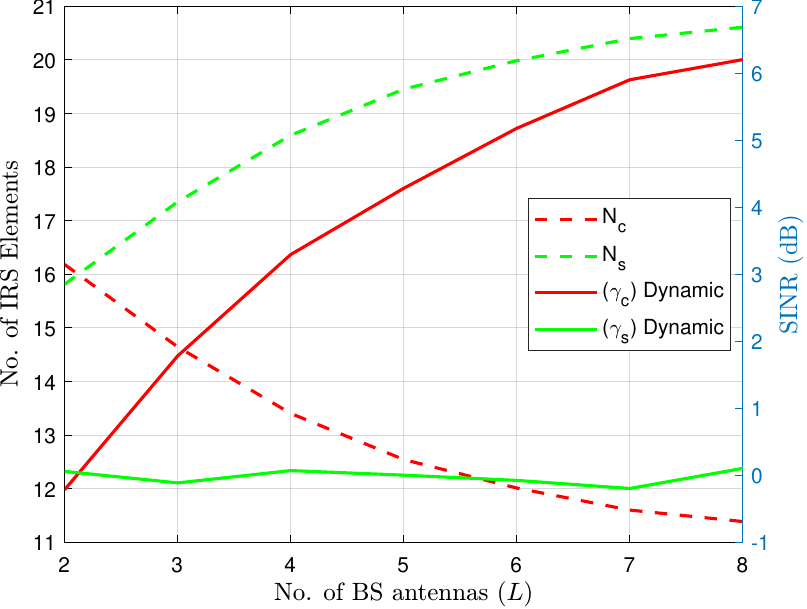}
    \caption{Effect of SINRs on the allocation of IRS elements with varying number of BS antennas.}
    \label{fig3:ALLOC}
\end{figure}

Dynamic element allocation plays a crucial role in this IRS-assisted spectrum-sharing environment, serving as the key distinction from previously employed multiple IRS schemes designed to assist communication and sensing. Fig. \ref{fig3:ALLOC} shows how element allocation dynamically responds to the calculated SINR values. SINRs $\gamma_c$ and $\gamma_s$ are plotted on the right y-axis of the figure against the varying number of BS antennas and corresponding element allocation is plotted on the left y-axis. Initially, when the SINR values of both systems are almost equal, the proposed scheme equally allocates the IRS elements to both systems. However, as the number of BS antennas increases, it intuitively increases the SINR of communication, and the gap between the two SINRs increases. Since element allocation weight ($\eta$) is the ratio of the two SINRs, more elements will be allocated to sensing. Another key observation is that the sensing SINR does not degrade as the number of BS antennas increases, despite increased interference at the radar. This is attributed to the regularized beamforming design, which effectively mitigates interference, along with the dynamic increase in the number of elements allocated to sensing.

\begin{figure}[h]
    \centering
    \includegraphics[width=0.42\textwidth]{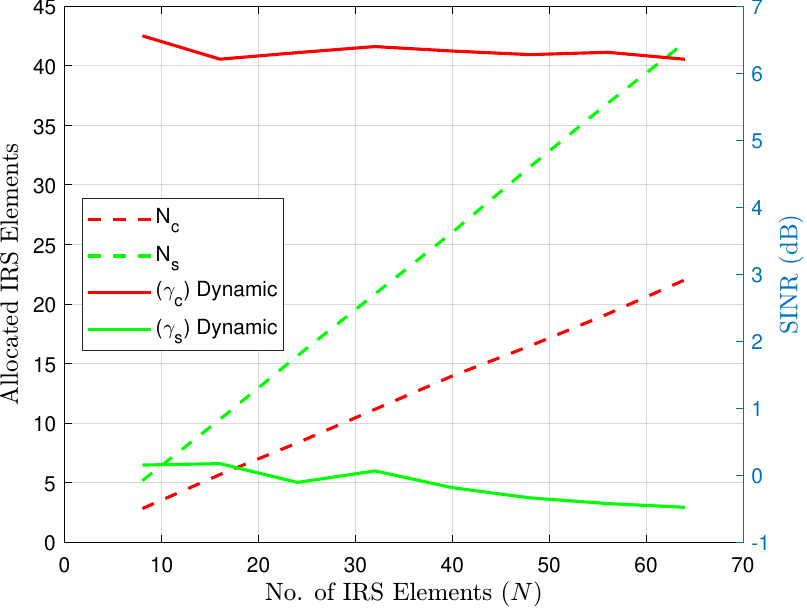}
    \caption{Effect of SINRs on the allocation of IRS elements with varying number of total IRS elements.}
    \label{fig4:NIRS}
\end{figure}

Finally, we present the effect of the total number of IRS elements on the SINR and dynamic element allocation in Fig. \ref{fig4:NIRS}. As the total number of IRS elements increases, the communication SINR remains stable however the sensing SINR faces some degradation. This is attributed to the increased interference because of the increasing number of allocated elements for communication. Dynamic allocation again comes into play an important role by adjusting the allocation to maintain a balance between the SINRs of the two systems.

\section{Conclusion}
\label{sec:Con}
We have presented an IRS-assisted spectrum-sharing framework for a communication and sensing network with a MIMO radar and a MIMO BS. Optimal performance for both the systems can be achieved by a robust interference mitigation strategy. The dynamic element allocation scheme plays a key role in improving the SINRs of both systems. Optimal beamforming and phase shifts serve as critical enablers for the dynamic element allocation strategy, ensuring an efficient balance between communication and sensing performance. Incorporating cross-interference into the design of beamforming and phase shifts is essential for accurate SINR calculations, which form the basis of dynamic allocation. Results show that the proposed optimization framework with a single IRS can perform better as compared to multiple IRS employed with fixed allocation for communication and sensing. The impact of imperfect CSI and a decentralized strategy for element allocation and interference mitigation is left as a topic for future work, which is currently underway.

\section*{Acknowledgments}
This work was supported in part by the German Federal Ministry for Education and Research (BMBF) within the projects Open6GHub \{16KISK004\}, and in part by Institute of Information \& communications Technology Planning \& Evaluation (IITP) grant funded by the Korea government (MSIT) (RS-2024-00397216).

\bibliographystyle{IEEEtran} 
\bibliography{reference.bib}


\end{document}